\documentclass[pre,twocolumn,floatfix,superscriptaddress,aps]{revtex4}
\usepackage{amsmath}
\usepackage{makeidx}
\usepackage{amssymb}
\usepackage{graphicx}

\usepackage[usenames]{color}
\usepackage[normalem]{ulem}

\newcommand{\beq}{\begin{equation}}
\newcommand{\eeq}{\end{equation}} 
\newcommand{\beqa}{\begin{eqnarray}}
\newcommand{\eeqa}{\end{eqnarray}} 
\begin{document}

\title{Spatially-periodic states  in a strongly dipolar $^{164}$Dy-$^{162}$Dy mixture}

\author{ S. K. Adhikari\footnote{sk.adhikari@unesp.br\\ 
https://professores.ift.unesp.br/sk.adhikari/}}
\affiliation{Instituto de F\'{\i}sica Te\'orica, UNESP - Universidade Estadual Paulista, \\ 01.140-070 S\~ao Paulo, S\~ao Paulo, Brazil}

\begin{abstract}

We demonstrate the formation of a novel  eigenstate   in a  strongly dipolar  binary $^{164}$Dy-$^{162}$Dy   mixture,  where the inter- and intraspecies dipolar lengths are larger than the corresponding scattering lengths. When this mixture is confined by a  quasi-two-dimensional harmonic trap,  the total density exhibits the formation of droplets on a spatially-symmetric triangular or square  lattice, where  each droplet is formed of a single species of atoms; two types of atoms never exist on the same lattice site.  The density of any of the species shows a partially-filled incomplete lattice, only the total density exhibits a completely full  lattice  structure.
  In this theoretical investigation  we employ the numerical solution of an improved
mean-field model including a Lee-Huang-Yang-type interaction in the intraspecies components alone, meant to stop a collapse of the atoms  at high
atom density.

\end{abstract}


\maketitle

\section{Introduction}

The experimental observation of a  Bose-Einstein condensate (BEC)
 of $^{52}$Cr \cite{ExpCr,cr,cr1,saddle,crrev,52Cr},  $^{166}$Er \cite{Er16},  
  $^{168}$Er \cite{ExpEr},
  $^{164}$Dy \cite{ExpDy,dy,dy1,dy2}, and  $^{162}$Dy  \cite{d249,d190}   atoms,
with large magnetic dipole
moment,  opened  a new avenue of research in search of  any new physics emerging from 
the anisotropic long-range dipolar interaction. 
Recently, two types of spatially symmetric eigenstates have been identified in a strongly dipolar BEC. 
The first is the experimental observation of a droplet of size much smaller than the harmonic oscillator trap lengths  \cite{drop1,drop2} in a harmonically-trapped strongly dipolar BEC of $^{164}$Dy atoms.
As  the number of atoms in this system is increased,   
multiple droplets arranged on a spatially-symmetric triangular  or a linear  lattice  were formed in a strongly dipolar BEC of $^{164}$Dy \cite{Dy11,Dy12}, $^{162}$Dy \cite{Dy13,Dy14,Dy15}, and $^{166}$Er \cite{Dy11,Dy12} atoms revealing  \cite{tri21} their  
 supersolid \cite{1,2,3,4,5,6} properties.  A supersolid or a solid superfluid, is a quantum
state of matter simultaneously possessing the properties
of both a solid and a superfluid,  where a spontaneous density modulation and a global phase coherence coexist. A supersolid has a spatially-periodic crystalline structure as a solid, breaking continuous translational invariance, and also enjoys
friction-less flow like a superfluid, breaking continuous
gauge invariance. In a trapped quasi-two-dimensional (quasi-2D)
dipolar BEC, the formation of a honeycomb lattice, stripe,
square lattice and other periodic and non-periodic patterns in density, were
also found  \cite{34,35,39,40,41,th3} in theoretical studies.

In view of the above-mentioned spatially-periodic states of distinct symmetry in a strongly dipolar BEC and the recent experimental observation of the binary dipolar  $^{164}$Dy-$^{166}$Er mixture\cite{dyer} and other studies on binary dipolar mixtures \cite{dyer1,dyer2,dyer3,dyer4,dyer5,dyer6,dyer7,dyer8,dyer9,dyer10} , in this paper we study the generation of similar spatially-periodic  arrangement of droplets on a triangular and a square lattice in a strongly dipolar binary 
$^{164}$Dy-$^{162}$Dy mixture.  
{ The droplet and droplet-lattice states are easily formed for a smaller number of atoms in a      $^{162}$Dy  BEC than in a $^{166}$Er BEC due to the larger dipole moment of the $^{162}$Dy atoms (10 Bohr magneton) compared to the  $^{166}$Er atoms (7 Bohr magneton).  For the same number of atoms, the net dipolar interaction in a Dy condensate is larger than that in a Er condensate, which is the key ingradient for the formation of droplets.} 
The strength of the dipolar interaction is measured by the dipolar length $a_{\mathrm{dd}}$, viz. Eq. (\ref{dl}), in the same  way as the scattering length determines the strength of the contact interaction in the mean-field model.
In the binary $^{164}$Dy-$^{162}$Dy mixture
 the interspecies and intraspecies dipolar lengths $a_{\mathrm{dd}}$ are approximately equal to $130a_0$ \cite{expt}, where $a_0$ is the Bohr radius.
We take the interspecies and intraspecies scattering lengths $a$ to be equal to $80a_0$.   This value of the scattering length, reasonably close to the experimental scattering length $a=92a_0$ \cite{expt,tang}, can be fixed by the use of the Feshbach resonance technique \cite{fesh} and    is found to be the most appropriate for the appearance of droplets for a small number of atoms.   
Under appropriate conditions again we find the formation of droplets and droplet lattice in each of the components in a quasi-2D harmonic trap. 
Due to the interspecies contact repulsion the droplets of one component stay in  distinct positions from the droplets of the other component; no droplet can have atoms from both components.  As a result,   a novel droplet lattice is formed, where each site can be occupied by atoms of the first or the second component. In this case the complete droplet-lattice structure appears in the total density; the component densities exhibit an incomplete droplet-lattice structure with  vacant sites. 

When the dipolar  interaction  is much stronger than the  contact interaction ($a_{\mathrm{dd}}>a$), as in this investigation, 
a dipolar mean-field Gross-Pitaevskii (GP)  equation \cite{crrev} cannot account for  the formation of a dipolar droplet or a droplet-lattice supersolid in a   strongly dipolar BEC  due to a collapse instability \cite{52Cr,collapse}. However, 
an improved \cite{improved} mean-field model including a higher-order quartic  Lee-Huang-Yang (LHY) type 
 \cite{lhy} repulsive interaction, appropriately modified due to the  dipolar interaction \cite{qf1,qf2,qf3}, can stop the collapse  and   explain   the formation of a dipolar droplet and  a  droplet-lattice  supersolid. We extend this improved mean-field model to the case of the binary dipolar-dipolar $^{164}$Dy-$^{162}$Dy mixture, which we use in the present study.   However, we introduce the LHY interaction only in  the Hamiltonian of the  intraspecies atoms of the two components which is sufficient for this investigation.  The inclusion of the LHY interaction in the  interspecies channel will just increase the repulsion in that channel and this should not have any qualitative effect on the   results presented in this paper. { The intraspecies repulsive LHY interactions alone stop the collapse and allow the formation of droplets. We verified that the addition of an interapecies repulsive interaction does not modify the scenario of the droplet formation and  will not be considered in this study. }

 We present results of a few complete triangular- and square-lattice supersolids for a small number  of atoms
 for  the binary $^{164}$Dy-$^{162}$Dy mixture, e.g. all possible permutations of the seven-droplet triangular-lattice state, 
and a few examples  of  the nineteen-droplet triangular-lattice state.  In addition, we present a few examples of the nine-droplet square-lattice state, and a few examples of the  twenty-five-droplet square-lattice states.  We consider states with an equal number of atoms in the two species as well as with different number of atoms in the two species.

In Sec. \ref{II}  we present the  improved binary mean-field model for a dipolar-dipolar mixture  including the repulsive LHY interaction in both intraspecies  dipolar components, which we use  in this numerical investigation.
The results of numerical calculation are illustrated in Sec. III, where  we detail the results for the six-droplet, seven-droplet, nine-droplet, nineteen-droplet and twenty-five-droplet  droplet-lattice states.  Of these the seven- and the nineteen-droplet states lie on a triangular lattice, whereas the nine- and the twenty-five-droplet states form a square lattice.   
Finally, in Sec. IV we present a brief summary of our findings.

\section{Improved mean-field model for a dipolar-dipolar mixture}

\label{II}

We consider a binary dipolar-dipolar BEC, interacting via  interspecies and intraspecies contact and dipolar interactions.  The 
mass, number of atoms, magnetic dipole moment,  scattering length, and dipolar length  for the two species, denoted by  $ i=1$, 2,
are given by $m_i, N_i, 
\mu_i, a_i, a_{\mathrm{dd}i}$, respectively.  However, in the $^{164}$Dy-$^{162}$Dy mixture, the magnetic dipole moments of the two species of atoms are identical, and consequently,    $\mu_1=\mu_2\equiv \mu$ = 10 Bohr magneton.
The  intraspecies ($V_i$) and 
interspecies ($V_{12}$)
interactions 
for two atoms  at positions $\bf r$ and $\bf r'$ are given by  \cite{crrev,expt}
\begin{eqnarray}\label{intrapot} 
V_1({\bf R})&=&
\frac{\mu_0\mu^2}{4\pi}V_{\mathrm{dd}}({\mathbf R})+\frac{4\pi 
\hbar^2 a_1}{m_1}\delta({\bf R }),
\end{eqnarray}
\begin{eqnarray}
V_2({\bf R})&=& \frac{\mu_0\mu^2}{4\pi}V_{\mathrm{dd}}({\mathbf R})+
\frac{4\pi 
\hbar^2 a_2}{m_2}\delta({\bf R }),\\
 \label{interpot} 
V_{12}({\bf R})&=& \frac{\mu_0\mu^2}{4\pi}V_{\mathrm{dd}}({\mathbf R})+
\frac{2\pi \hbar^2 a_{12}}{m_R}\delta({\bf R}),\\ \label{dp}
V_{\mathrm{dd}}({\mathbf R})  &=& 
\frac{1-3\cos^2\theta}{{\mathbf R}^3},
     \end{eqnarray}
where $\bf R \equiv (r-r')$ is the position vector joining the two atoms, $\mu_0$ is the permeability of free space, 
$\theta$ is the angle made by the vector ${\bf R}$ with the polarization 
$z$ direction,   $m_R=m_1m_2/(m_1+m_2)$ is the reduced mass of the two species of 
atoms, and $a_{12}$  is the interspecies  scattering length. 
To compare the dipolar and contact interactions, the intraspecies and interspecies 
dipolar interactions  are  expressed in terms of the dipolar lengths, defined by 
\begin{align}
  a_{\mathrm{dd}i}\equiv 
\frac{\mu_0\mu^2m_i}{12\pi \hbar ^2},  \quad  a_{\mathrm{dd}12}\equiv 
\frac{\mu_0\mu^2  2m_R}{12\pi \hbar ^2}.
\end{align}
   The effect of the corresponding intraspecies contact interaction is quantized by the scattering length $a_i$. 
The dimensionless relative dipolar length, defined by, 
\begin{equation} \label{dl}
\varepsilon_{\mathrm{dd}i}     \equiv \frac{a_{\mathrm{dd}i}}{a_i}  
\end{equation}
gives
the strength of the dipolar interaction relative to  the contact interaction  
and is useful to study  many properties of a dipolar BEC.

The angular frequencies for the axially-symmetric quasi-2D harmonic trap on both species of dipolar atoms
along $x$, $y$ and $z$ directions are taken as 
$\omega_x=\omega_y\equiv \omega_\rho$   ($\boldsymbol \rho =\{x,y\},$  $\rho^2=x^2+y^2,$) and 
$\omega_z (\gg \omega_\rho)$.  
 With intraspecies and  interspecies interactions (\ref{intrapot})-(\ref{interpot}), the  improved coupled   GP
equations for the binary dipolar-dipolar  BEC mixture can be written as \cite{mfb,mfb1,mfb2} 
\begin{align}
\label{eq1}
{\mbox i} \hbar \frac{\partial \psi_1({\bf r},t)}{\partial t}  &=
{\Big [}  -\frac{\hbar^2}{2m_1}\nabla^2+ \frac{1}{2}m_1  
(\omega_\rho ^2\rho^2+\omega_z^2 z^2 )
\nonumber\\ &
+ \frac{4\pi \hbar^2}{m_1}{a}_1 N_1 \vert \psi_1({\bf r},t) \vert^2
+\frac{2\pi \hbar^2}{m_R} {a}_{12} N_2 \vert \psi_2({\bf r},t) \vert^2
\nonumber 
\\  
&+\frac{3\hbar^2}{m_1}a_{\mathrm{dd}1} N_1 \int  V_{\mathrm{dd}} ({\mathbf R})\vert\psi_1({\mathbf r'},t)\vert^2 d{\mathbf r}' \nonumber \\
&+\frac{3\hbar^2}{2m_R}a_{\mathrm{dd}12} N_2 \int  V_{\mathrm{dd}} ({\mathbf R})\vert\psi_2
({\mathbf r'},t)\vert^2 d{\mathbf r}' \nonumber \\
&+\frac{\gamma_{\mathrm{LHY}1}\hbar^2}{m_1}N_1^{3/2}
|\psi_1({\mathbf r},t)|^3
\Big] 
 \psi_1({\bf r},t),
\\
{\mbox i} \hbar \frac{\partial \psi_2({\bf r},t)}{\partial t}  &=
{\Big [}  -\frac{\hbar^2}{2m_2}\nabla^2+ \frac{1}{2}m_2  
(\omega_\rho ^2\rho^2+\omega_z^2 z^2 )
\nonumber\\ &
+ \frac{4\pi \hbar^2}{m_2}{a}_2 N_2 \vert \psi_2({\bf r},t) \vert^2
+\frac{2\pi \hbar^2}{m_R} {a}_{12} N_1 \vert \psi_1({\bf r},t) \vert^2
\nonumber 
\\  
&+\frac{3\hbar^2}{m_2}a_{\mathrm{dd}2} N_2 \int  V_{\mathrm{dd}} ({\mathbf R})\vert\psi_2({\mathbf r'},t)\vert^2 d{\mathbf r}' \nonumber \\
&+\frac{3\hbar^2}{2m_R}a_{\mathrm{dd}12} N_1 \int  V_{\mathrm{dd}} ({\mathbf R})\vert\psi_1({\mathbf r'},t)\vert^2 d{\mathbf r}' \nonumber \\
&+\frac{\gamma_{\mathrm{LHY}2}\hbar^2}{m_2}N_2^{3/2}
|\psi_2({\mathbf r},t)|^3
\Big] 
 \psi_2({\bf r},t),
\label{eq2}
\end{align}
where ${\mbox i}=\sqrt{-1}$ and we have included the LHY interaction only in the intraspecies channels which is sufficient for our purpose to stop the collapse.  
The    LHY interaction  coefficients  $\gamma_{\mathrm{LHY}i}$ \cite{lhy} in  Eqs. (\ref{eq1}) and (\ref{eq2}) are given by \cite{qf1,qf3,qf2}
\begin{align}\label{qf}
\gamma_{\mathrm{LHY}i}= \frac{128}{3}\sqrt{\pi a^5} Q_5(\varepsilon_{\mathrm{dd}i}), 
\end{align}
where  the auxiliary function $ Q_5(\varepsilon_{\mathrm{dd}i})$ includes the  
correction to the       LHY interaction due to the dipolar interaction  in the intraspecies channels \cite{qf1,qf2}  and is
 given by \cite{arxiv}
\begin{align}
 Q_5(s)&=\ (1-s)^{5/2}  {_2F_1} \left(-\frac{5}{2},\frac{1}{2};\frac{3}{2};\frac{3s}{s-1}\right)\\ &
 \equiv\int_0^1 du(1-s+3su^2 )^{5/2},  
\end{align}
where we have used an integral representation of the hyper-geometric function  $_2F_1$.  The auxiliary function $Q_5$ 
{ can be written as   \cite{th1,blakie}
\begin{align}\label{exa}
Q_5(s) &=\
\frac{(3s)^{5/2}}{48}   \Re \left[(8+26\eta+33\eta^2)\sqrt{1+\eta}\right.\nonumber\\
& + \left.
\ 15\eta^3 \mathrm{ln} \left( \frac{1+\sqrt{1+\eta}}{\sqrt{\eta}}\right)  \right], \quad   \eta = \frac{1-s}{3s},
\end{align}
where $\Re$ is the real part.   { 
 The perturbative result (\ref{qf}) is  valid for weakly dipolar atoms 
($\varepsilon_{\mathrm{dd}i}<1$), while $Q_5(\varepsilon_{\mathrm{dd}i})$ is real  \cite{qf1,qf2}.  However,  in this study of strongly dipolar atoms $s\equiv \varepsilon_{\mathrm{dd}i}= a_{\mathrm{dd}i}/a\approx 1.6 >1$. 
 Indeed, for $\varepsilon_{\mathrm{dd}i}> 1,$ 
the dipolar interaction  dominates over the repulsive contact interaction leading to the collapse of the condensate and the function $Q_5(\varepsilon_{\mathrm{dd}i})$ becomes imaginary \cite{qf3}.
The imaginary part of $Q_5(s)$  is not just a mathematical peculiarity,  it implies the collapse
instability of the strongly dipolar system \cite{qf3}.
  In this case of dysprosium atoms, it has been shown  that the imaginary part of  $Q_5(\varepsilon_{\mathrm{dd}i})$  is small \cite{young},  which is neglected here and elsewhere \cite{expt,blakie,saddle,Ex2}. 
  For $s= 0$ and 1, the expression (\ref{exa}) is indeterminate
and 
$Q_5 (0) = 1$ and $Q_5(1) = 3 \sqrt 3/2$ [40].
In this paper we use expression (\ref{exa}) for $Q_5(s)$.
}

 Equations  (\ref{eq1}) and (\ref{eq2})  can be written in the following dimensionless form if we scale lengths in units of 
 $l = \sqrt{\hbar/m_1\omega_z}$,
time in units of $\omega_z^{-1}$,    angular frequency in units of $\omega_z$,
energy in units of $\hbar \omega_z$ and density $|\psi_i|^2$   
in units of $l^{-3}$ 
\cite{mfb,mfb1,mfb2,th1}
\begin{align}
\label{eq3}
{\mbox i}  \frac{\partial \psi_1({\bf r},t)}{\partial t}  &=
{\Big [}  -\frac{1}{2}\nabla^2
+ \frac{1}{2}(\omega_\rho ^2\rho^2+ z^2 )
+ g_1 \vert \psi_1({\bf r},t) \vert^2
\nonumber \\
&+g_{12}\vert \psi_2({\bf r},t) \vert^2 
+ g_{\mathrm{dd}1}      \int  V_{\mathrm{dd}} ({\mathbf R})\vert\psi_1({\mathbf r'},t)\vert^2 d{\mathbf r}' \nonumber \\
&+ g_{\mathrm{dd}12}      \int  V_{\mathrm{dd}} ({\mathbf R})\vert\psi_2({\mathbf r'},t)\vert^2 d{\mathbf r}' \nonumber \\
&+\gamma_{\mathrm{LHY}1}N_1^{3/2}
|\psi_1({\mathbf r},t)|^3
\Big] 
 \psi_1({\bf r},t),
\end{align}
\begin{align}
{\mbox i}  \frac{\partial \psi_2({\bf r},t)}{\partial t} &=
{\Big [}  -\frac{m_{12}}{2}\nabla^2
+ \frac{m_{21}}{2}(\omega_\rho ^2\rho^2+ z^2 )
+ g_2 \vert \psi_2({\bf r},t)\vert^2
\nonumber
\\  & 
+g_{21} \vert \psi_1({\bf r},t)|^2 
+ g_{\mathrm{dd}2}      \int  V_{\mathrm{dd}} ({\mathbf R})\vert\psi_2({\mathbf r'},t)\vert^2 d{\mathbf r}' \nonumber \\
&+ g_{\mathrm{dd}21}      \int  V_{\mathrm{dd}} ({\mathbf R})\vert\psi_1({\mathbf r'},t)\vert^2 d{\mathbf r}' \nonumber \\
&+m_{12}\gamma_{\mathrm{LHY}2}N_2^{3/2}
|\psi_2({\mathbf r},t)|^3
{\Big ]}  \psi_2({     \bf r},t),
\label{eq4}
\end{align}
where
$m_{12}={m_1}/{m_2},
m_{21}={m_2}/{m_1},
g_1=4\pi a_1 N_1,
g_2= 4\pi a_2 N_2 m_{12},
g_{12}={2\pi m_1} a_{12} N_2/m_R,
g_{21}={2\pi m_1} a_{12} N_1/m_R,
g_{\mathrm{dd}1}= 3N_1 a_{\mathrm{dd}1}, g_{\mathrm{dd}2}= 3N_2 a_{\mathrm{dd}2}m_{12},g_{\mathrm{dd}12}= 3N_2 a_{\mathrm{dd}12}m_1/2m_R, g_{\mathrm{dd}21}= 3N_1 a_{\mathrm{dd}12}m_1/2m_R.$
 
For a stationary state, Eqs. (\ref{eq3}) and (\ref{eq4}) can be obtained from a minimization 
of the following energy functional  (total energy of all atoms in the two components)
\begin{align}\label{energyx}
E&= \frac{1}{2}\int d{\bf r}\Big[N_1|\nabla \psi_1({\bf r})|^2 
+m_{12} N_2|\nabla \psi_2({\bf r})|^2   
\nonumber \\
&+N_1(\omega_\rho^2\rho^2+ z^2)|\psi_1({\bf r})|^2+N_2m_{21}(\omega_\rho^2\rho^2+ z^2)|\psi_2({\bf r})|^2
\nonumber \\&
+\sum_i {N_i}g_i|\psi_i({\bf r})|^4  +2N_1g_{12}|\psi_1({\bf r})|^2   |\psi_2({\bf r})|^2 
\nonumber \\
&+\sum_i {N_i}g_{\text{dd}i}\int d{\bf r}' V_{\text{dd}}({\bf R})|\psi_i({\bf r'})|^2   |\psi_i({\bf r})|^2 
\nonumber \\
&+2{N_1}g_{\text{dd}12}\int d{\bf r}'  V_{\text{dd}}({\bf R})|\psi_1({\bf r'})|^2   |\psi_2({\bf r})|^2
\nonumber \\
&+\frac{4}{5}\gamma_{\mathrm{LHY}1}N_1^{5/2} |\psi_1({\bf r})|^5+ \frac{4}{5}m_{12}\gamma_{\mathrm{LHY}2}N_1^{5/2} |\psi_2({\bf r})|^5
\Big] 
\end{align}
 via the rule
 \begin{align}
 \mbox{i}\frac{\partial \psi_i}{\partial t}= \frac{\delta E}{\delta \psi_i^ *} .
 \end{align}

\section{Numerical Results} 

\label{III}

\begin{figure*}[!t]

\begin{center}

\includegraphics[width=\linewidth]{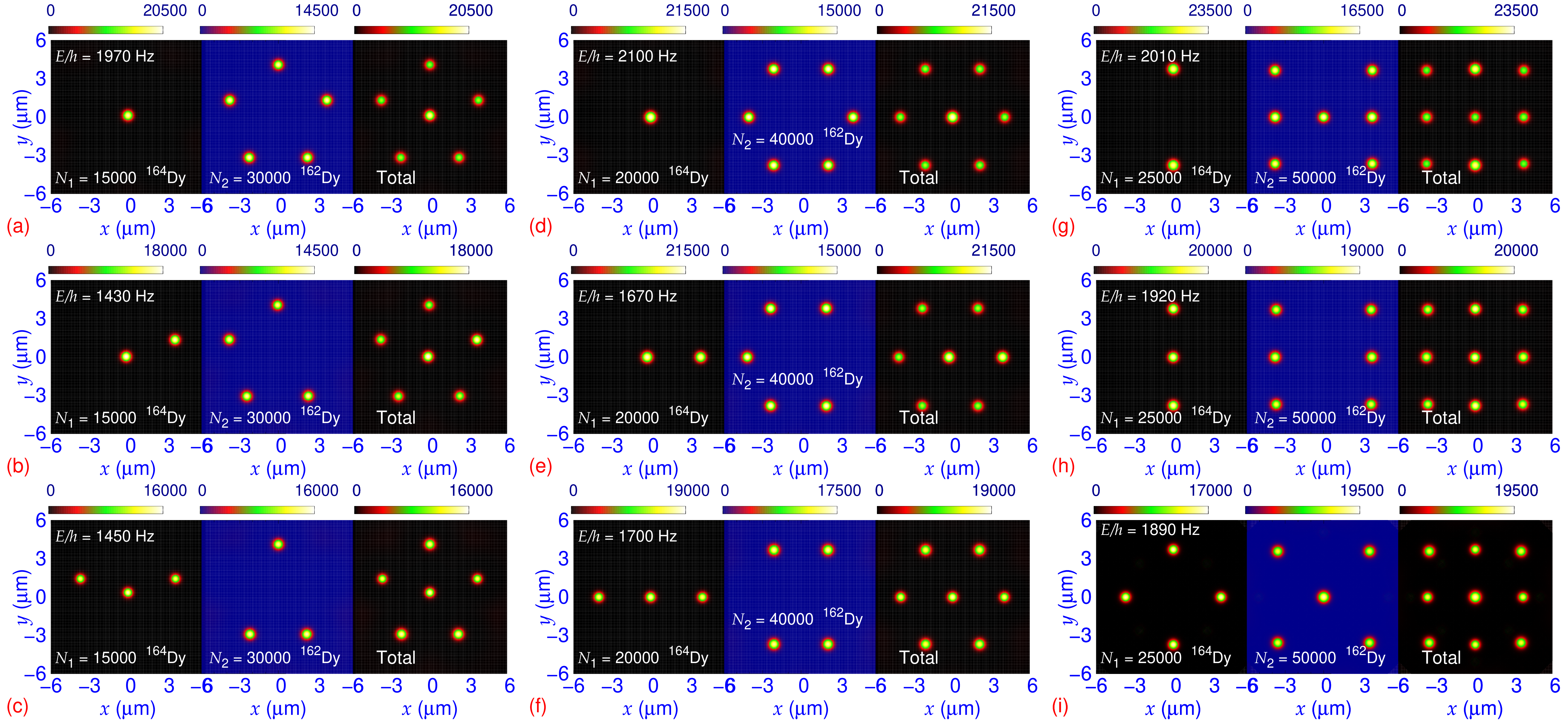} 
\caption{ (Color online)  Contour plot of quasi-2D densities $n_i(x,y)$ in units of $\mu$m$^{-2}$  of  
 $^{164}$Dy atoms ($i=1$, left panel),  $^{162}$Dy atoms ($i=2$, middle panel) and total density 
 [$n_1(x,y)+n_2(x,y)$, right panel] in the $^{164}$Dy-$^{162}$Dy mixture.  In (a)-(c) a few possible 
 six-droplet  states of $N_1=15000$ $^{164}$Dy atoms and $N_2=30000$ $^{162}$Dy atoms are illustrated.
 In (d)-(f) [(g)-(i)] a few possible 
 seven-droplet [nine-droplet]  states of $N_1=20000$  [$N_1=25000$] $^{164}$Dy atoms and $N_2=40000$ [$N_2=50000$] $^{162}$Dy atoms are presented.
       The intraspecies and interspecies scattering lengths are all taken to be equal: 
 $a_1=a_2=a_{12}=80a_0$; the dipolar lengths are $a_{\mathrm{dd}1}(^{164}$Dy)=130.8$a_0$,  $a_{\mathrm{dd}2}(^{162}$Dy)=129.3$a_0$,  $a_{\mathrm{dd}12}(^{164}$Dy-$^{162}$Dy)=130.0$a_0$.  
Both species of atoms are confined by  an  axially-symmetric quasi-2D trap with angular frequencies $\omega _z=2\pi \times 167$ Hz,  $\omega _ x=\omega _y=2\pi \times 33$ Hz.    
 }
     \label{fig1}
\end{center}
\end{figure*}

To study the eigenstates of the dipolar-dipolar $^{164}$Dy-$^{162}$Dy mixture, the partial differential GP equations (\ref{eq3}) and (\ref{eq4}) are solved, numerically, using C/FORTRAN programs \cite{49} or their open-multi-processing versions \cite{52,53}, using the split-time-step Crank-Nicolson method. 
We will use imaginary-time propagation to find the stationary eigenstates employing a space step of {$0.1l$ and a time step $0.001\times \omega_z^{-1}$} \cite{Santos01,CPC,CPC1}.  The solution obtained by imaginary-time propagation is the same as that obtained by a minimization of energy (\ref{energyx}).
It is difficult to evaluate numerically  the divergent $1/|{\bf R}|^3$ term in the dipolar potential (\ref{dp})
in configuration space.
The contribution of the nonlocal dipolar interaction integral in coupled Eqs. (\ref{eq3}) and (\ref{eq4}) is calculated numerically in the momentum 
space by a fast Fourier transformation routine  using a convolution theorem \cite{Santos01,49}.  The Fourier transformation of the dipolar interaction in the  momentum space  is analytically known \cite{49}.   In this way the problem is solved
in the momentum space and the solution in configuration
space is obtained by taking another Fourier transformation.

 The intraspecies  and interspecies dipolar lengths of  $^{164}$Dy and  $^{162}$Dy atoms  are   $a_{\mathrm{dd}1}=130.8a_0,$ $a_{\mathrm{dd}2}=129.3a_0$ and   $a_{\mathrm{dd}12}=130.0 a_0$    \cite{expt}.
 The angular  frequencies of the   quasi-2D  axially-symmetric trap used in this paper  to study  droplet-lattice states are taken as 
 $\omega_\rho\equiv \omega_x=\omega_y=2\pi \times  33$ Hz, 
$\omega_z=2\pi \times 167$ Hz, as in some recent experiments on a quasi-2D  supersolid formation with $^{164}$Dy 
atoms
 \cite{tri21,Ex2} and also used in some theoretical investigations \cite{th1,Th2,39}.
{    In this study $m_1 = m(^{164}$Dy) $=164 \times 1.66054 \times 10^{-27}$ kg, consequently, the unit of length $l=\sqrt{\hbar /m_1\omega_z} = 0.607$ $\mu$m. The unit of time is $\omega_z ^{-1}=0.953$ ms.   In imaginary-time propagation the employed space step is 0.0607 $\mu$m and time step is 0.953 $\mu$s.                   
}

 The formation of the droplet-lattice 
 is best confirmed by a consideration of the integrated reduced quasi-2D density 
 $n_i(x,y)$ defined by \cite{ajp}
 \begin{equation}
 n_i(x,y)=N_i\int_{-\infty}^{\infty}  dz|\psi_i(x,y,z)|^2,
 \end{equation}
as well as the total density of the two components $n_1(x,y)+n_2(x,y)$.



In a numerical calculation by imaginary-time propagation, to find a specific droplet-lattice state easily,  the choice of the initial state is essential. { To obtain the droplet-lattice states in a controlled way in the binary system, we first generated, by imaginary-time propagation following Ref. \cite{39}, a desired droplet-lattice state in the single-component dysprosium condensate  in the same harmonic trap and for the same scattering and dipolar lengths.  The so-obtained converged state, after certain modifications, is used as the initial state of both  components in the present study of the binary $^{164}$Dy-$^{162}$Dy mixture. For example, in the calculation of a seven-droplet state on a triangular lattice in the binary mixture, if we would like  three specific droplet-lattice sites to  be occupied by the first component, viz. Fig. \ref{fig1}(f), in the initial state of the first component we take only these three droplets; in the first component  the remaining droplet-lattice sites are left vacant.   The initial state of the second component is left unchanged with all lattice sites filled with droplets. After imaginary-time propagation with these initial states, the first component remains qualitatively unchanged with the three droplets in the desired sites. Due to interspecies repulsion, during imaginary-time propagation, the atoms of the droplets of the second component moves away from the occupied sites of the first component and eventually the second component  forms  droplets only at the unoccupied  sites of the first component.  In this fashion, the desired supersolid of the binary mixure has droplets of both the components.  }    This procedure is adopted in all calculations of this paper to find a specific stationary state and obtain it energy.

The formation of the present droplet-lattice state is best illustrated through a contour plot of the quasi-2D density,
$n_i(x,y), i=1,2$ in the $x$-$y$ plane,
 of each of the components as well as the  total  density $[n_1(x,y)+n_2(x,y)]$
 and we illustrate these in Fig. \ref{fig1} for a few of the six-,  seven- and nine-droplet states.  In Figs.\ref{fig1}(a)-(c)  [(d)-(f), (g)-(i)] we display the contour plots of the quasi-2D densities of a few of the six-droplet 
[seven-droplet, nine-droplet]
states  of  $N_1=15000 $  [$N_1= 20000, 25000$]  $^{164}$Dy atoms   and  $N_2=30000$  [$N_2=40000,50000$] $^{162}$Dy atoms  and of the total  density.  No complete lattice structure can be built with the six-droplet state, but the seven-droplet (nine-droplet) state fills  a triangular (square) lattice.
As we target a larger number of droplets in the states of Figs. \ref{fig1}(d)-(i) compared to those in  Figs. \ref{fig1}(a)-(c), the number of atoms in the components has been increased successively for an efficient formation of droplets. 
In all cases, $N_2= 2N_1$. Hence in the lower-energy states the number of the droplets in the first component is usually smaller than the number of droplets in the second component as shown in Fig. \ref{fig1}, where we show a few of the lower-energy states. { The states with a larger number of droplets in the first component are excited states  with larger energy. Such states may not be obtainable by imaginary-time propagation as imaginary-time propagation is designed to converge to the lower-energy states, although, this method may find some low-energy excited states with distinct symmetry.}  Hence we display in Fig. \ref{fig1} only the states with 
$N_{\mathrm{drop}2} \ge  N_{\mathrm{drop}1}$, where $N_{\mathrm{drop}i}$ is the number of droplets in component $i$.
 The same is true about the states with a very small number of droplets in the first component, viz. the states in Figs. \ref{fig1}(a), (d), and (g). For $N_1=25000$ and $N_2=50000$ the state with one droplet in the first component and eight droplets in the second component is unstable. The state with two droplets in the first component is metastable as presented in Fig. \ref{fig1}(g).  It is not easy to find the stable ground state for a given value of $N_1$ and $N_2$, because there could be  different permutations of droplets in a state even if the number of droplets in the components are fixed.  For example, corresponding to the state of Fig. \ref{fig1}(a)  there is another distinct state with one droplet in the first component, where $^{164}$Dy atoms occupy one of the outer sites and not the central site as in this plot. The same is true about the state of Fig. \ref{fig1}(d). However, these metastable states have a larger energy than the states of Figs. \ref{fig1}(a) and (d) (results not shown in this paper). The states in Figs. \ref{fig1}(a)-(f)  are the lowest-energy states for one, two, and three droplets in the first component. Of these states, those in Figs. \ref{fig1}(b) and (e)  with $N_{\mathrm{drop}2} >  N_{\mathrm{drop}1}$ are the ground states among the states with total number of six and seven droplets, respectively.  
Corresponding to the state of Fig. \ref{fig1}(b) [\ref{fig1}(e)]  there are two  [three]   more distinct states with two droplets in the first component. 
Corresponding to the state of Fig. \ref{fig1}(c) [\ref{fig1}(f)]  there are two  [three]   more distinct states with three droplets in the first component. Figures \ref{fig1}(a)-(c)  [Figs. \ref{fig1}(d)-(f)] are the minimum energy six-droplet [seven-droplet] states with 1, 2, 3 droplets in the first component.  The number of possible permutations in case of the nine droplet states in Figs. \ref{fig1}(g)-(i) is large and we did not do a numerical analysis to find the minimum-energy states.  
The states illustrated in Figs. \ref{fig1}(g)-(i) are just examples of nine-droplet states. 
 In Fig. \ref{fig1}
the density of each of the components show a partial occupation of the underlying  lattice. The total density of the two components reveals the spatially symmetric triangular- and square-lattice  structure in Figs. \ref{fig1}(d)-(f) and \ref{fig1}(g)-(i), respectively.   Each of the components fill in complementary sites.

 \begin{figure}[!t]

\begin{center}
 
\includegraphics[width=\linewidth]{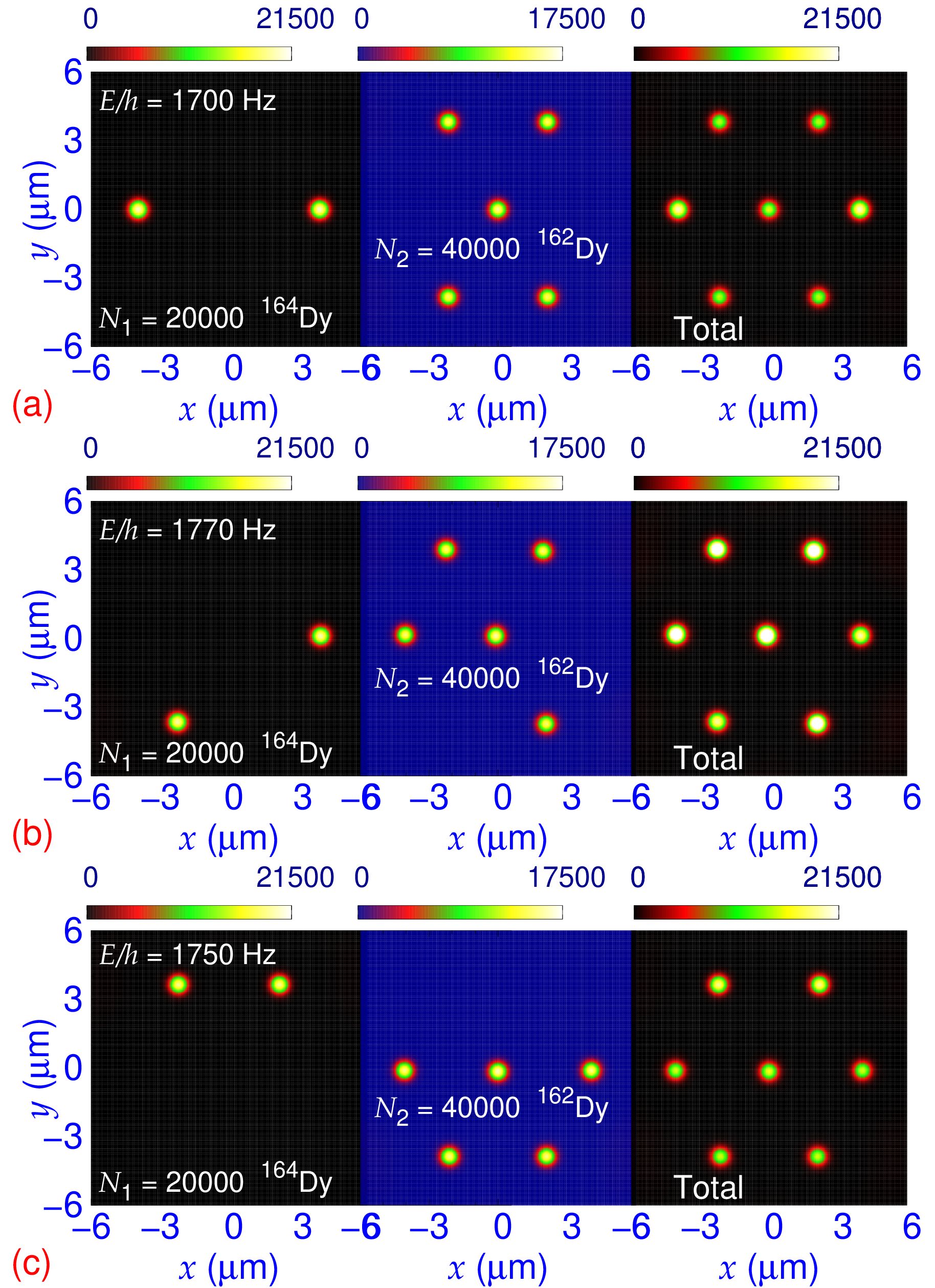}

\caption{ (Color online) Contour plot of quasi-2D density of a few seven-droplet  triangular-lattice states  in units of $\mu$m$^{-2}$  of  
 $^{164}$Dy atoms (left panel),  $^{162}$Dy atoms (middle panel) and total density 
 (right panel) in the $^{164}$Dy-$^{162}$Dy mixture with two (five) droplets in component 1 (2).  
The number of atoms in the two components are  $N_1=20000$, $N_2=40000$.  All other parameters in this calculation are the same as in Fig. \ref{fig1}.
  }\label{fig2}
\end{center}
\end{figure}
   
 \begin{figure}[!t]  

\begin{center}
 
\includegraphics[width=\linewidth]{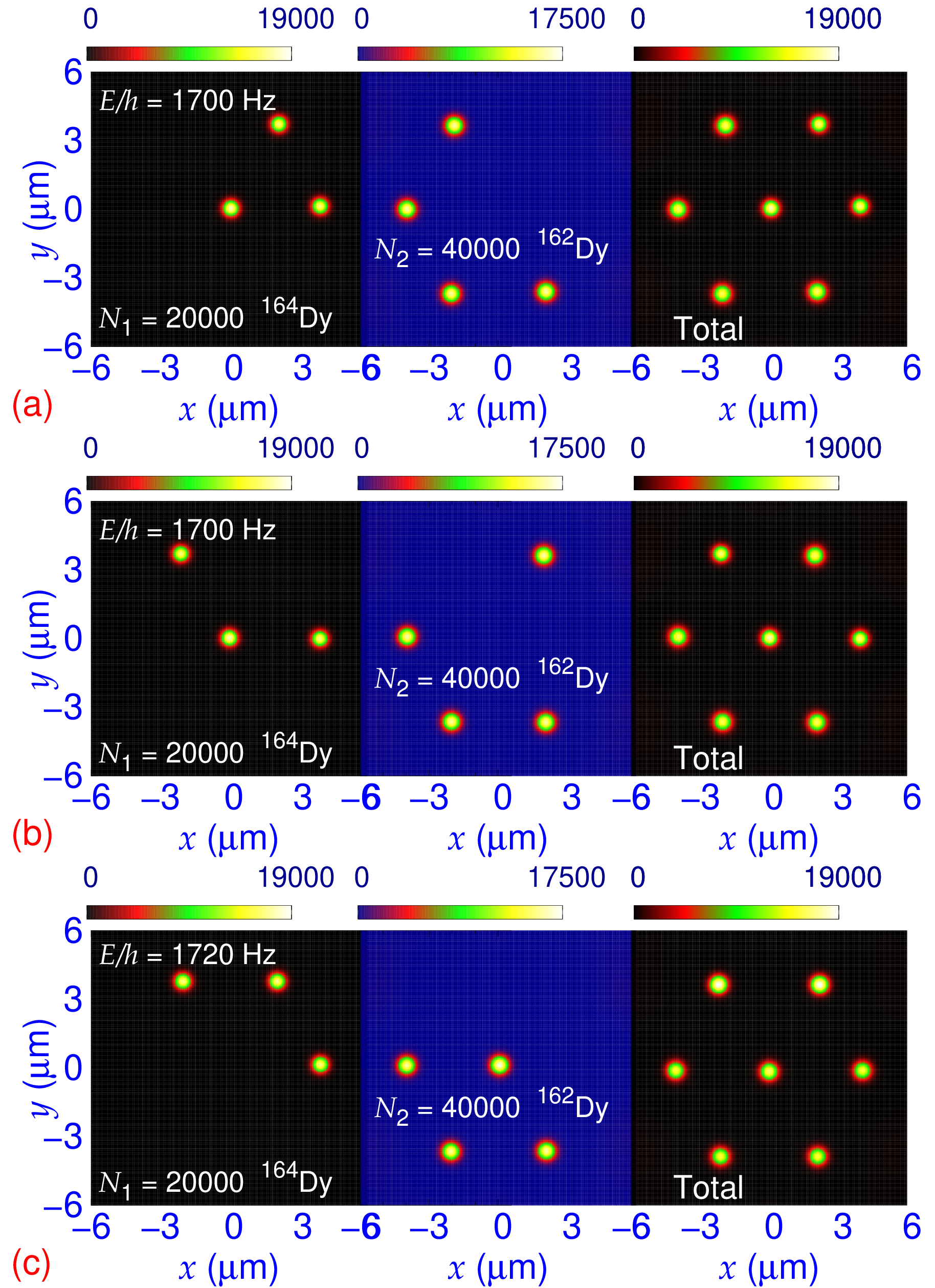}

\caption{ (Color online) Contour plot of quasi-2D density of a few seven-droplet triangular-lattice states  in units of $\mu$m$^{-2}$  of  
 $^{164}$Dy atoms (left panel),  $^{162}$Dy atoms (middle panel) and total density 
 (right panel) in the $^{164}$Dy-$^{162}$Dy mixture with three (four) droplets in component 1 (2).  
The number of atoms in the two components are  $N_1=20000$, $N_2=40000$.  All other parameters in this calculation are the same as in Fig. \ref{fig1}.
  }\label{fig3}
\end{center}
\end{figure}

  As the triangular lattice is the most common 2D lattice and as it is also found in experiments on a strongly dipolar BEC \cite{drop1,drop2,tri21}, we find it pertinent to investigate the different distinct metastable permutations   of the states in Figs. \ref{fig1}(e) and (f) with two and three droplets in the first component and five and four droplets in the second component, respectively. In Fig. \ref{fig2} we illustrate the distinct permutations of the ground state displayed  in Fig. \ref{fig1}(e). All these states have an energy larger than the same of the state in Fig. \ref{fig1}(e). The  different permutations of the state in Fig. \ref{fig1}(f) with three droplets in component 1 and four in component 2 are illustrated in Fig. \ref{fig3}.  From a consideration of energy of these states, we find that the states in Figs. \ref{fig1}(f) and \ref{fig3}(a)-(b) are {approximately degenerate as verified numerically}, but the ground state of all the seven droplet states is the one in Fig. \ref{fig1}(e) with two droplets in the first component and seven droplets in the second component.

\begin{figure}[!t]

\begin{center}
 
\includegraphics[width=.95\linewidth]{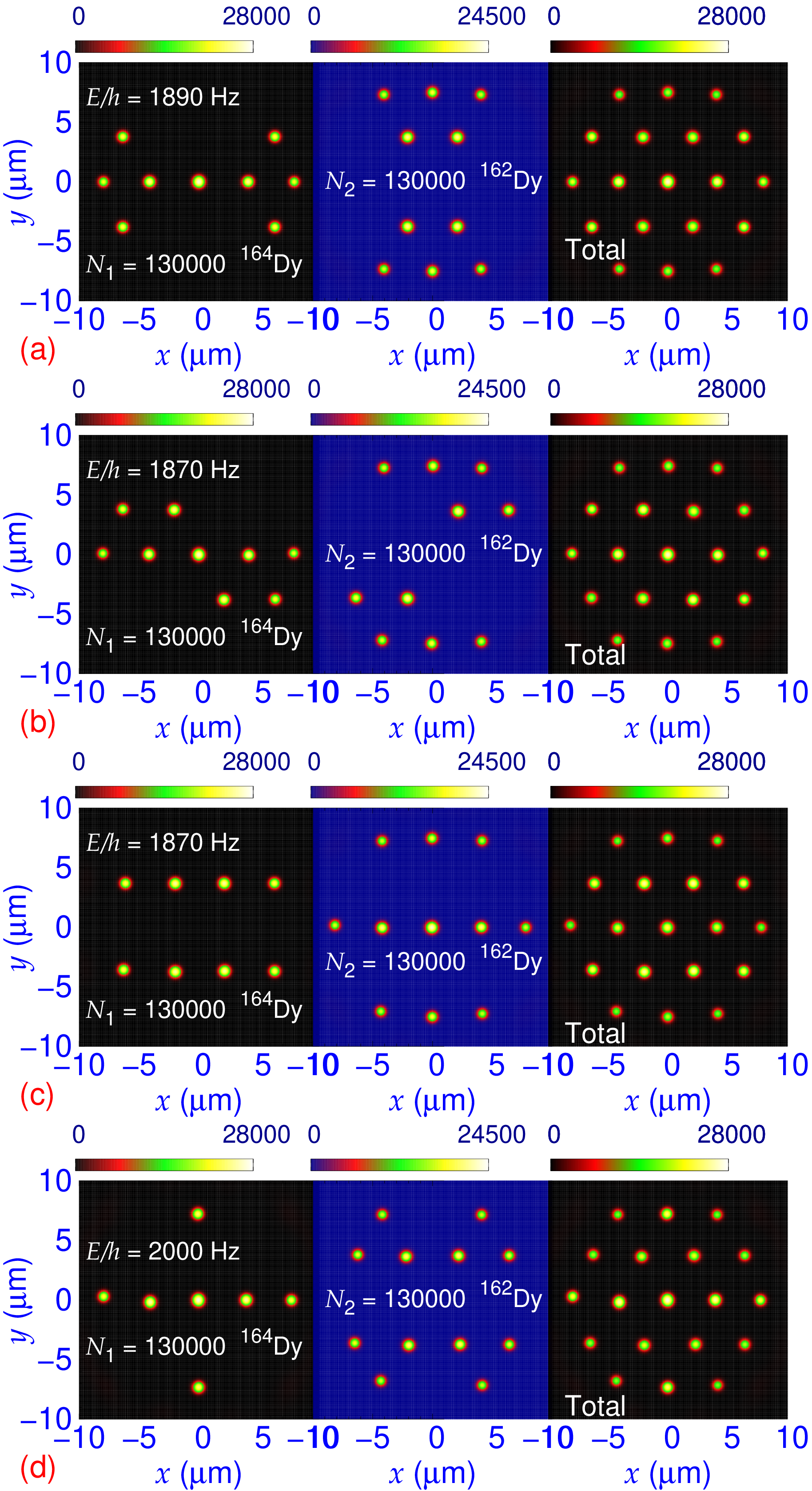}

\caption{ (Color online) Contour plot of quasi-2D density of a few nineteen-droplet triangular-lattice states  in units of $\mu$m$^{-2}$  of  
 $^{164}$Dy atoms (left panel),  $^{162}$Dy atoms (middle panel) and total density 
 (right panel) in the $^{164}$Dy-$^{162}$Dy mixture with (a)-(b) nine  (ten) droplets in the first (second) component, (c) eight (eleven) droplets in the first (second) component, and (d) seven (twelve) droplets     in the first (second) component.
The number of atoms in the two components are  $N_1=N_2=130000$.  All other parameters in this calculation are the same as in Fig. \ref{fig1}.
  }\label{fig4}
\end{center} 
\end{figure}

\begin{figure}[!t]

\begin{center}
 
\includegraphics[width=.95\linewidth]{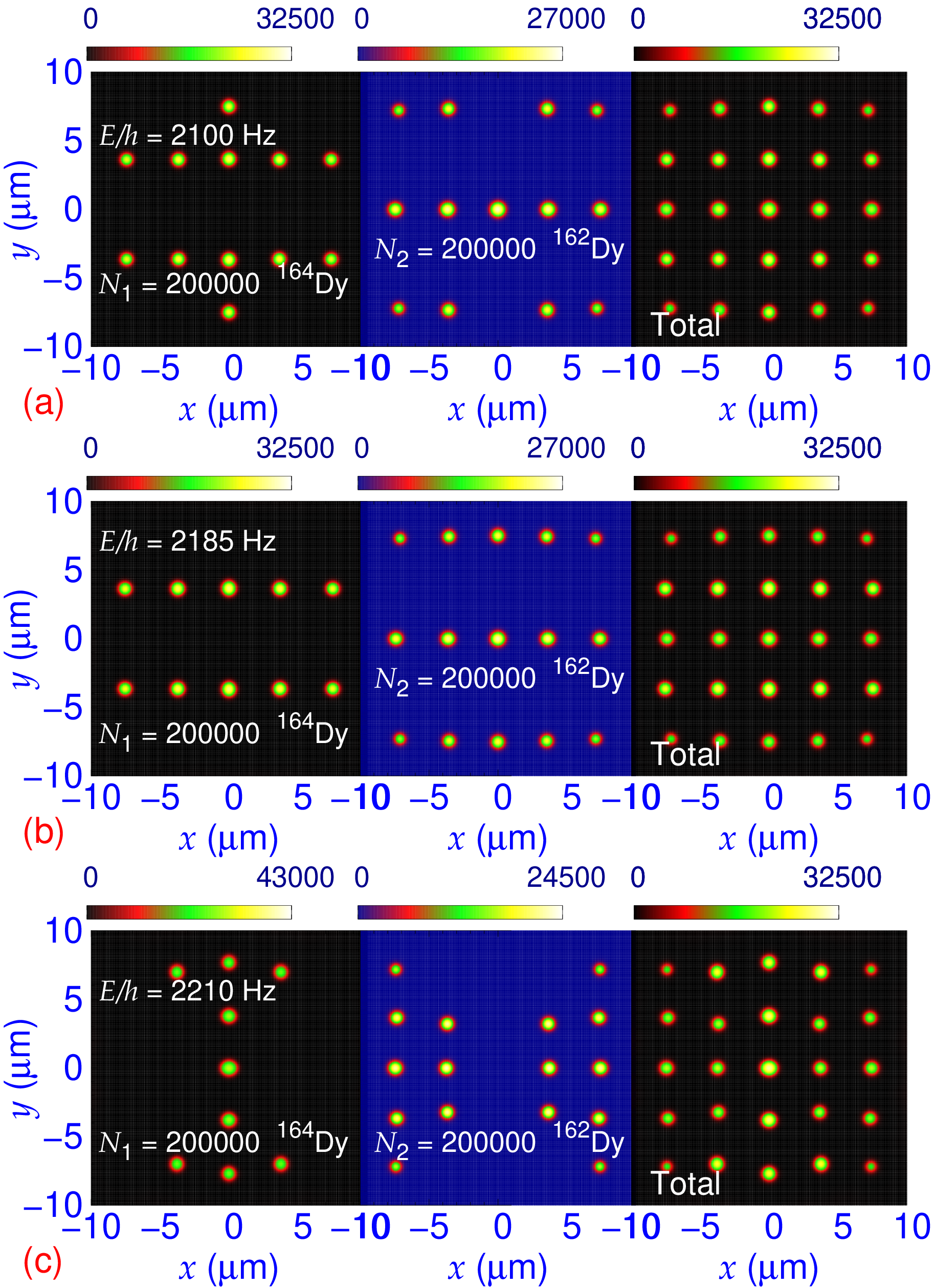}

\caption{ (Color online)  Contour plot of quasi-2D density of a few twenty-five-droplet square-lattice states  in units of $\mu$m$^{-2}$  of  
 $^{164}$Dy atoms (left panel),  $^{162}$Dy atoms (middle panel) and total density 
 (right panel) in the $^{164}$Dy-$^{162}$Dy mixture with (a) twelve  (thirteen) droplets in the first (second) component, (b) eight (fifteen) droplets in the first (second) component, and (c) nine (sixteen) droplets     in the first (second) component.
The number of atoms in the two components are  $N_1=N_2=200000$.  All other parameters in this calculation are the same as in Fig. \ref{fig1}.
  }\label{fig5}
\end{center} 
\end{figure}

Next we show a few examples of triangular and square lattice states with larger number of atoms, so that a triangular (square) droplet-lattice of nineteen (twenty five) droplets is completely filled in.  In this case we take the number of atoms in the two components to be equal $N_1=N_2$.  After a small experimentation, the number of atoms in case of a triangular  (square) lattice is taken as $N_1=N_2=130000$ ($N_1=N_2=200000$). In this case it is not possible to make a comprehensive study of all possible states as the number of possible permutations is very large. As the triangular lattice has been experimentally observed we give several examples of this case.
In Figs. \ref{fig4}(a) we display  a nineteen-droplet triangular-lattice state where the first (second) component has nine  (ten) droplets through a contour plot of quasi-2D densities $n_i(x,y)$ of $^{164}$Dy (first component) and $^{162}$Dy
(second component)  versus $x$ and $y$.   In Fig \ref{fig4}(b) we show a different permutation of this state with a slightly different energy.  In these cases the number of droplets in the two components are approximately equal (9 and 10 droplets), although the number of atoms are the same ($N_1=N_2=130000$).  It is also possible to have a state with a larger disbalance in the number of droplets in the two components  maintaining all parameters unchanged.  In Fig. \ref{fig4}(c) we illustrate a state with eight droplets in the first component and eleven droplets in the second component. In this case the alternate rows  are filled in by droplets of different species. In Fig. \ref{fig4}(d)  we present a state with seven droplets 
in the first component and twelve droplets in the second component.  It has a much larger energy than  the previous states indicating that it is less probable:  nature prefers a state with similar number of droplets in the two components, as the number of atoms in the two components are equal.

Finally, in Fig. \ref{fig5} we present a  few possible square-lattice twenty-five-droplet state with $N_1=N_2=200000$ atoms in each of the components.  In this case both species have the same number of atoms and the most probable  state(s)  should also have similar number of droplets $-$ twelve (thirteen) in  the first (second) component $-$ as shown in Fig. \ref{fig5}(a).  In Figs. \ref{fig5}(b)-(c) we illustrate two states with a larger imbalance of the number of droplets in the two components, e.g. ten and nine in  the first component and fifteen and sixteen in the second component, respectively.  These states are less probable as the number of atoms in the two components are equal, and hence, these states have larger energy than the state in Fig. \ref{fig5}(a) with a similar number of droplets in the two components.

\section{Summary and Discussion}

Using a numerical solution of a   set of improved binary mean-field GP equations,  we study states of a strongly dipolar binary mixture $^{164}$Dy-$^{162}$Dy  and demonstrate a new type of formation of droplets on square and triangular lattices.  In this study  the interspecies and intraspecies scattering lengths and dipolar lengths   are taken as $ a_i=a_{12}=80a_0,  a_{\mathrm{dd}i}=a_{\mathrm{dd}12}
\approx  130a_0, i=1,2$, respectively, which makes the system strongly  dipolar and appropriate for an efficient formation of droplets.  
The complete lattice structure is  found in the binary mixture  and each component of the mixture leads to an  incomplete lattice structure.  Each droplet is formed by a single species of atom and the interspecies repulsion excludes the possibility of a  single lattice site being occupied by two types of atoms. 
For this calculation we employ imaginary-time propagation, which preserves the symmetry of the  arrangement of the droplets in each component. { In the case of a usual GP equation in the weak-coupling limit, the imaginary-time propagation is designed to converge to the parity-symmetric ground state. However, if a parity-antisymmetric initial state is employed this method converges to the first excited parity-antisymmetric state. Hence the imaginary-time propagation usually preserves the symmetry of the initial state. In the present complex problem of binary dipolar mixture, it was possible to obtain the excited states with complex symmetry properties by imaginary-time propagation by employing an appropriate initial state.}  For a triangular lattice, we consider completely filled in states with seven and nineteen droplets    and for a square lattice we consider also completely filled in  states with nine and twenty five droplets.
In both cases we consider equal and different number of atoms in the two species. 
In the case of equal number of atoms in the two species the ground  state favors one with approximately equal number of droplets in the two species.  
The present study may open a new avenue of theoretical and  experimental  research  on the formation of supersolid states in strongly dipolar binary mixture. In view of the experimental observation of a strongly dipolar  $^{164}$Dy-$^{166}$Er  
mixture \cite{dyer},  the experimental observation of $^{164}$Dy-$^{162}$Dy  seems possible in order to investigate the findings of this paper.

\acknowledgments

SKA acknowledges support by the CNPq (Brazil) grant 301324/2019-0.

\end{document}